\documentclass[twocolumn,aps,amsmath]{revtex4}
\usepackage[dvips]{graphics}
\newcommand{\be}{\begin{equation}}
\newcommand{\ee}{\end{equation}}

\newcommand{\bea}{\begin{eqnarray}}
\newcommand{\eea}{\end{eqnarray}}
\newcommand{\bd}{\begin{displaymath}}
\newcommand{\ed}{\end{displaymath}}
\newcommand{\bi}{\begin{itemize}}
\newcommand{\ei}{\end{itemize}}
\newcommand{\bc}{\begin{center}}
\newcommand{\ec}{\end{center}}
\newcommand{\bfl}{\begin{flushleft}}
\newcommand{\efl}{\end{flushleft}}
\newcommand{\bfr}{\begin{flushright}}
\newcommand{\efr}{\end{flushright}}
\newcommand{\f}{\frac}

\def\6{\partial}  
  \def\ve{\varepsilon}

\def\={\!\!\!&=&\!\!\!}
\def\+{\!\!\!&&\!\!\!+~}
\def\-{\!\!\!&&\!\!\!-~}

\begin{document}
\title{Excitonic condensation in quasi-two-dimensional systems}
\author{M. Crisan}
\affiliation{Department of Theoretical Physics, University of
Cluj, 400084 Cluj-Napoca, Romania}
\author{I. \c{T}ifrea}
\affiliation{Department of Theoretical Physics, University of
Cluj, 400084 Cluj-Napoca, Romania}
\affiliation{Department of Physics and Astronomy, The University
of Iowa, Iowa City, IA 52242, USA}

\begin{abstract}
We present a low energy  model for the Bose-Einstein condensation
in a quasi-two-dimensional excitonic gas. Using the flow equations
of the Renormalization group and a $\Phi^4$ model with the
dynamical critical exponent $z=2$ we calculate the temperature
dependence of the critical density, coherence length, magnetic
susceptibility, and specific heat. The model can be relevant for
the macroscopic coherence observed in  GaAs/AlGaAs  coupled
quantum wells.
\end{abstract}

\maketitle

\section{Introduction}

The Bose--Einstein condensation (BEC) in interacting bosonic
systems has attracted increasing attention in the recent years due
to the practical realization of a condensate state in trapped
alkali-metal atoms. Also, a sustained experimental effort was
directed to the realization and visualization of the BEC of
excitons in low dimensional systems, namely in semiconductor
quantum wells (QW).\cite{butov,snoke} A remarkable property of the
BEC state is that a large number of particles, which are in the
condensate state, become dependent on a single wave function
promoting quantum properties to classical length and time scale.
This state can be described by a wave function with a phase
coherence over distances much larger than separation between
individual particles. Along with BEC states in pure bosonic
systems such as alkali atoms, there are a series of BEC states
formed from composite bosons. For example, such a state appears in
superconductors, where the Cooper pairs can be treated as
composite bosons formed from two electrons which interact via an
attractive potential. In standard superconductors the average
characteristic length of the Cooper pairs (the coherence length)
is much larger than the actual distance between pairs, making the
BEC state a state of overlapping Cooper pairs. This situation is
in contrast with the case of a BEC state in pure bosonic systems
such as helium and atomic vapors of metals, where the bosons are
single particles and the system can be treated in a dilute limit.

The BEC state in semiconductors is obtained from the condensation
of excitons, which are also composite bosons formed from an
electron bounded to a hole. Excitons are usually created by
shining light on the semiconductor, as a result an equal number of
electrons and holes being created. The electrons interact with the
holes leading to the formation of electron--hole (e--h) bound
states. Theoretically the formation of excitons was predicted to
appear in semiconductors\cite{keld} or metals.\cite{over} In
metallic systems such bound states lead to the itinerant electron
antiferromagnetism studied initially in the mean field
approximation.\cite{feders,rice,mircea} On the other hand, in
semiconductor bulk systems, the short life time of optically
generated excitons was a major obstacle to the experimental study
of such a system. The recent advances in the realization of low
dimensional semiconductor systems such as QW made possible to
overcome the issues of a short life time in bulk systems.
Semiconductor QW allow the confinement of electrons and holes in
layered configurations, which can be assimilated to quasi-two
dimensional (2D) systems.

A slightly different situation occurs in the case of bilayer QW
where indirect excitons can be formed between the conduction band
electrons from one layer and the valence band holes  from the
adjacent layer. The spatial separation between electrons and holes
give rise to a repulsive interaction  between the excitons, which
prevent the occurrence of the e-h plasma. Experimental studies of
the bilayer QW system showed an enhance exciton mobility, an
increased radiative decay rate, and a photoluminescence noise. All
these results are considered to be an indirect argument for the
BEC of the excitons as they are connected to an increased
coherence among the system. A detailed and clear discussion of the
particular features of the BEC in bosonic systems formed by
fermions (Cooper pairs and excitons) was given a long time ago by
Kohn and Sherrington.\cite{kohn} The critical behavior and the
role of excitonic fluctuations have been investigated using the
renormalization group (RNG) method,\cite{baba} with the main
result being the evidence of a dimensional crossover between a
semi-metallic and a semiconducting state in three dimensional (3D)
systems. Such a crossover effect is determined by the band overlap
and is influenced by the presence of non-magnetic impurities in
the system.

Here, we will study using the RNG the possibility of a BEC of
excitons in systems with a reduced dimensionality. In such
systems, which usually are considered to be quasi-2D systems, the
BEC is known as a quasi-condensation because it does not appear in
the thermodynamic limit. The model we explore is similar to the
one applied to a 2D bosonic system,\cite{crisan} used to address
the physical properties of the condensed state in He. Based on
this model we investigate the thermodynamic properties of the
system as the temperature dependence of the critical density,
coherence length, magnetic susceptibility, and specific heat.

\section{Renormalization group approach}

The bosonic system formed by the excitons can be described by the
following general action:
\begin{equation}\label{eq1}
S=\frac{1}{2}\int d\tau d^d{\bf
r}\left[\Phi^*\6_\tau\Phi-(\nabla^2-\mu)|\Phi|^2+\frac{t_0}{8}|\Phi|^4\right]\;,
\end{equation}
where $\Phi\equiv\Phi(\bf r,\tau)$ is the bosonic field describing
the $e-h$ density fluctuations, $t_0$ is the bare interaction
between the bosons and $\mu$ is the effective chemical potential.
In the case of an excitonic gas formed in a semiconductor system
the effective chemical potential is defined as:
\begin{equation}
\mu(n)=\varepsilon_g -\varepsilon_0\;,
\end{equation}
where $\varepsilon_g$ is the semiconducting band gap and
$\varepsilon_0$ is the exciton binding energy. The spatial scale
of the problem, for which the interaction between the component
excitons treated as bosons becomes relevant, is determined by the
smallest length in the system, namely, the radius of the exciton,
$r_0\sim\ve_0^{1/2}$. The existence of a low density parameter is
determined by the range $r_0$ of the interaction potential, $t_0$,
and the mean separation between the constituent particles,
$\bar{l}$ ($\mu\sim1/(\bar{l}^2)$). In this case
\be
r_0\sqrt{\mu}\ll 1\;.
\ee
Under this condition we can describe the condensation of the
excitons in a large temperature interval, including the critical
region.

To address the phase transition in the excitonic gas we used the
method introduced by Popov \cite{popov} to describe the standard
Bose gas. Accordingly, we introduce an intermediate momentum $p_i$
such that $\mu<\ve_i<\ve_0$, with $\ve_i=p_i^2/(2m)$. The
effective interaction between the constituen particles, $u_0$, is
obtained in the t-matrix approximation as:
\begin{equation}\label{uzero}
u_0=\frac{t_0}{1+t_0\Pi}\;.
\end{equation}
In Eq. (\ref{uzero}) the polarization, $\Pi$, is given by:
\begin{equation}\label{polarizare}
\Pi=\int^{p_i}_{1/r_0}\frac{d^d k}{(2\pi)^d}\frac{1}{k^2}.
\end{equation}
The integration in Eq. (\ref{polarizare}) is strongly dependent on
the system dimensionality. For the case of a two-dimensional
system, $d=2$, the effective interaction can be approximated as
\begin{equation}\label{uzero2D}
u_0\simeq\frac{2}{\pi\ln\frac{\varepsilon_0}{\varepsilon_i}}\;,
\end{equation}
a value which according to the energy scale of the problem is
small. On the other hand, for $d>2$, the effective interaction can
be calculated as $u_0\simeq C r_0^  {(d-2)}$, where $C$ is a
constant. For the two dimensional case, the polarization is
logarithmical small, and the resulting effective interaction is
repulsive and small. As a result we can use the RNG formalism,
applied to a modified action, to describe the Bose gas formed by
the constituent excitons. The resulting action may describe a
quantum phase transition (QPT) in the two dimensional exciton Bose
system at finite temperature. The formalism is similar to that
developed in Refs. \onlinecite{crisan,ileana} for the weakly
interacting Bose system.

The standard Wilson RNG procedure, leads to the following set of
differential equations corresponding to the renormalized chemical
potential, $\mu(l)$, interaction, $u(l)$, and temperature, $T(l)$:
\begin{equation}\label{renmu}
\frac{d\mu(l)}{dl}=2\mu(l)+K_2 F_1(r,T)u(l)\;,
\end{equation}
\begin{equation}\label{renu}
\frac{du(l)}{dl}=-\frac{K_2}{4}u(l)^2\;,
\end{equation}
and
\begin{equation}\label{renT}
\frac{dT(l)}{dl}=2T(l)\;.
\end{equation}
Here $K_2=1/(2\pi)$ and $ F_1(\mu,T)$ is given by:
\begin{equation}
F_1(\mu,T)=\frac{1}{2}\coth\frac{1}{2T(l)}\;.
\end{equation}
This set of differential equations can be solved starting from the
last equation to the first one. From Eqs. (\ref{renT}) and
(\ref{renu}), we get
\begin{equation}\label{Tsol}
T(l)=T\exp(2l)
\end{equation}
and
\begin{equation}\label{usol}
u(l)=\frac{1}{C(l+l_0)}\;,
\end{equation}
with $C=K_2/4$ and $l_0=8\pi/u_0$. Following the same methods as
in Refs. \onlinecite{crisan,ileana} we calculated the renormalized
chemical potential, $\mu(l)$, as:
\bea\label{solmu}
\mu(l)&=&\exp[\Lambda(l)]\left[\mu-\mu_c-\f{1}{2\pi}\int_0^l\f{dl'
e^{-2l'}
u(l')}{e^{1/T(l')}-1}\right]\nonumber\\&&-\frac{u(l)}{8\pi}\;,
\eea
where $\mu_c=u_0/(8\pi)$ and $\Lambda(l)$ is given by the
expression:
\begin{equation}
\Lambda(l)=2l\left[1-\frac{1}{l}\ln\left(1+\f{l}{l_0}\right)\right]\;.
\end{equation}
The possibility of a phase transition in the excitonic system can
be studied if we introduce a new parameter, $t_\mu(l)$, defined
as:
\begin{equation}\label{tmu}
t_\mu(l)=\mu(l)+\frac{u(l)}{8\pi}\;.
\end{equation}
Based on Eqs. (\ref{usol}) and (\ref{solmu}), $t_\mu(l)$ can be
written as:
\begin{equation}\label{tmuexplicit}
t_\mu(l)=\exp{[\Lambda(l)]} t_\mu(T)\;,
\end{equation}
where
\begin{equation}
t_\mu(T)=\left[t_\mu(0)+\frac{u_0T}{4\pi}\int_a^\infty\frac{dx}{\exp{[x]}-1}\right]\;,
\end{equation}
with $a=u_0/(4\pi)$. The renormalization procedure will be stopped
at $l=l^*$, where $l^*$ is the solution of the following equation:
\begin{equation}\label{l*def}
t_\mu(l^\star)=1.
\end{equation}
From Eqs. (\ref{tmu}) and (\ref{tmuexplicit}) we find $l^\star$
as:
\begin{equation}\label{lstar}
\exp{(-2l^\star)}\cong\frac{T}{\ln(1/T)}\;,
\end{equation}
a value which in the following will be used to study the main
thermodynamic properties of the excitonic system.

\section{Thermodynamics in the critical region}

The RNG permits the evaluation of the main thermodynamical
properties of the system in the critical region around the QPT
point. The correlation length, $\xi$, and the magnetic
susceptibility, $\chi$, can be obtained using the procedure
presented in Refs. \cite{crisan,ileana} as:
\begin{equation}\label{xi}
\xi(\mu,u_0,T)\simeq\xi_0\exp(l^\star)
\end{equation}
and:
\begin{equation}\label{chi}
\chi(\mu,u_0,T)\simeq\chi_0\exp(2l^\star)\;,
\end{equation}
where $\xi_0$ and $\chi_0$ are constants and the value for
$l^\star$ given by the Eq. (\ref{lstar}). A simple calculation
leads to:
\begin{equation}\label{xifinal}
\xi(T)\simeq\xi_0\frac{|\ln{(1/T)}|^{1/2}}{T^{1/2}}
\end{equation}
and:
\begin{equation}\label{chifinal}
\chi(T)\simeq\chi_0\frac{|\ln{(1/T)}|}{T}\;.
\end{equation}
The critical density, $n(T)$, will be calculated using the
renormalized Bose-Einstein function considered in $l=l^*$:
\begin{equation}\label{density}
n(T)=\exp(-2l^{\star})\int_0^{\infty}\frac{d^2k}{(2\pi)^2}
\left\{\exp\left[\f{k^2+\mu}{T(l^*)}\right]
-1\right\}^{-1}\;.
\end{equation}
In the case of a two-dimensional system the integral can be
calculated analytically leading to
\begin{equation}\label{densityfinal}
n(T)\simeq T\ln{|\ln{(1/T)|}}\;.
\end{equation}
This equation can be inverted to calculate the critical
temperature for the BEC in the excitonic system \cite{crisan}:
\begin{equation}
T_c(n)\simeq\frac{n}{\ln{|\ln{(1/n)}|}}\;.
\end{equation}

Another important parameter of the critical region is the specific
heat. In order to obtain its temperature dependence we consider
the scaling of the free energy, $f(l)$, written as \cite{crisan}:
\begin{equation}\label{freeenergy}
\frac{df(l)}{dl}=4f(l)+g[\mu(l), T(l)]\;,
\end{equation}
where :
\begin{equation}\label{functiong}
g(l)\simeq \frac{K_2}{2}\int_0 ^l dl'\exp(-4l')[1+\mu(l')]\;.
\end{equation}
The specific heat in the critical region will be calculated using
the standard definition, $C(T)=-T\6^2f(T)/\6T^2$, using for the
chemical potential the renormalized value obtained from Eq.
(\ref{solmu}). A simple calculation leads to a temperature
dependence whose most divergent contribution is given by:
\begin{equation}\label{caldura}
 C(T)\simeq\frac{T}{|\ln{T}|^3}\;,
\end{equation}
a result similar to that obtained for the two-dimensional Bose
system \cite{crisan}.

\section{Conclusion}

In this paper we developed a formalism based on the RNG approach
for bosonic systems in order to describe the physical properties
of an excitonic gas. We consider the case of a two dimensional
system, a configuration which is very close to the quasi two
dimensional geometries of GaAs quantum wells, where an excitonic
condensation can be observed. Our result shows that in the case of
a two dimensional system the critical temperature for the BEC in
the excitonic system has a double logarithmical dependence on the
density. Similar studies \cite{ket} showed a different dependence
of the critical temperature. However, the models considered in
Ref. \cite{ket} are three dimensional models with an in-plane
harmonic or square potential confinement, which cannot reproduce
the exact result from the two dimensional calculations.

The many-body methods for the calculation of the optical
absorbtion-gain spectra has been formulated in Ref. \cite{chu} at
finite temperatures. The authors claim that fluctuations cannot
destroy the condensation of excitons even at finite temperature.
The effects of exciton BEC can be observed up to 130 K, which is a
very high temperature. However, it is well known that the
mean--field approximations overestimate the critical temperature,
but the results are important because they show how to include the
short--range Coulomb potential with screening in the problem.


Here we proposed a microscopic description of the excitonic
condensation in a two dimensional ($d=2$) system in terms of an
effective action similar to that for the interacting Bose liquid
with repulsive short--range interaction. For this model the
effective coupling constant between the component excitons was
evaluated using a t-matrix approximation, and based on the
characteristic energy scale we showed that it is small.
Accordingly, the RNG method in the one-loop approximation can give
relevant results. The critical density presents a non-linear
temperature dependence, which can describe the behavior of the
experimental results in a relevant temperature interval. A similar
non-linear dependence was observed experimentally \cite{butov}.
The critical temperature, compared to the expression obtained in
Ref. \cite{ket} and used in Ref. \cite{butov} to explain the
experimental behavior is much smaller. A more realistic model, has
to consider the influence of trapping in the RNG formalism for the
two dimensional system, but such a calculation is much more
complicated. However, even this oversimplified model showed the
importance of the quantum effects in the condensation of the
excitons.

\end{document}